\documentclass[pdflatex,sn-nature]{sn-jnl}
\AtBeginDocument{\fontsize{10pt}{14pt}\selectfont}
\usepackage{graphicx}%
\usepackage{multirow}%
\usepackage{amsmath,amssymb,amsfonts}%
\usepackage{amsthm}%
\usepackage{mathrsfs}%
\usepackage[title]{appendix}%
\usepackage{xcolor}%
\usepackage{textcomp}%
\usepackage{manyfoot}%
\usepackage{booktabs}%
\usepackage[detect-weight=true,detect-family=true]{siunitx}
\usepackage{algorithm}%
\usepackage{algorithmicx}%
\usepackage{algpseudocode}%
\usepackage{listings}%
\theoremstyle{thmstyleone}%
\theoremstyle{thmstyletwo}%
\theoremstyle{thmstylethree}%
\usepackage{fancyhdr}

\fancypagestyle{titlepage}{
    \fancyhf{}
    \fancyfoot[C]{%
        \small
        \begin{tabular}{c}
        $\dagger$ These authors contributed equally.\\
        *Corresponding authors. E-mails: shar0486@umn.edu; greven@umn.edu
        \end{tabular}%
    }

}

\begin{document}

\title{Piezomagnetism in a model cubic noncollinear altermagnet}
\author*[1]{\fnm{Sudarshan} \sur{Sharma}$^\dagger$}

\author[2]{\fnm{Luca} \sur{Buiarelli}$^\dagger$}

\author[1]{\fnm{Richard} \sur{Spieker}}

\author[3]{\fnm{Ivan} \sur{Jakovac}}

\author[1,3]{\fnm{Damjan} \sur{Pelc}}

\author[2]{\fnm{Turan} \sur{Birol}}

\author*[1]{\fnm{and Martin} \sur{Greven}}

\affil[1]{\orgdiv{School of Physics and Astronomy},
\orgname{University of Minnesota},
\orgaddress{\city{Minneapolis}, \state{MN}, \postcode{55455}, \country{USA}}}

\affil[2]{\orgdiv{Department of Chemical Engineering and Materials Science},
\orgname{University of Minnesota},
\orgaddress{\city{Minneapolis}, \state{MN}, \postcode{55455}, \country{USA}}}

\affil[3]{\orgdiv{Department of Physics},
\orgname{Faculty of Science},
\orgname{University of Zagreb},
\orgaddress{\city{Zagreb}, \country{Croatia}}}

\date{\today}

\abstract{\bf 
Altermagnets constitute a distinct class of magnetic materials that combine compensated magnetic order with spin-polarized electronic bands, hence carrying characteristics of both antiferromagnets and ferromagnets. Piezomagnetism -- the linear relationship between lattice strain and a net magnetic moment -- has emerged as smoking-gun evidence of altermagnetism that distinguishes it from antiferromagnetism. Here, we uncover a large piezomagnetic response in MnTe$_2$, a cubic altermagnet with a noncollinear spin arrangement and weak spin-orbit coupling. We combine dilatometry with nuclear magnetic resonance, a bulk local probe, to observe signatures of both direct and inverse piezomagnetism consistent with symmetry considerations and first-principles calculations. The dilatometry measurements reveal a shear deformation proportional to an applied magnetic field, while magnetic resonance detects a ferromagnetic moment induced by shear strain. The respective results for the piezomagnetic coupling strength are in good agreement, and they are captured by first-principles results. These findings for simple binary MnTe$_2$ position this material as a model altermagnet, demonstrate the utility of nuclear magnetic resonance in investigations of piezomagnetism, and open new avenues for multimodal strain and magnetic-field control in spintronic and memory applications.
}


\renewcommand{\abstractname}{}

\maketitle
\thispagestyle{titlepage}
\clearpage
Magnetically ordered materials without net magnetization have recently attracted renewed interest due to the emergence of new classes, such as altermagnets and odd-parity magnets \cite{SmejkalPRX2022, jungwirth2025altermagnetism, Yu2025odd}. These materials have real-space compensated magnetic configurations similar to conventional antiferromagnets, yet exhibit nodal splitting of spin-up and spin-down bands even without spin-orbit coupling~\cite{krempasky2024altermagnetic}. The defining feature of altermagnetic order is the breaking of combined space-time (\textit{PT}) symmetry, while the two opposite spin sublattices remain connected by a real-space rotation (proper or improper and symmorphic or nonsymmorphic) \cite{SmejkalPRX2022}. This has various consequences for the electronic properties, including chiral splitting of magnon bands \cite{LiuPRLChiralSpilting}, 
%
elasto-Hall conductivity \cite{elastohallconductvity}, as well as spin Hall and Edelstein effects \cite{hu2025spinhallncalm}, which make altermagnets suitable candidates for emerging technological applications \cite{jungwirth2026altermagnetic, fualtermagneticspintronics} and a fertile playground to study the interconnection between magnetism, symmetry, and novel electronic phenomena \cite{park_impact_2026, fang_quantum_2024, banerjee_altermagnetic_2024}. 

The prediction of nonrelativistic spin splittings in compounds without a net magnetization was followed by both theoretical \cite{haule2025, vandanbrink2023} and experimental claims \cite{krempasky2024altermagnetic, fedchenko2024observation} of altermagnetic order in a vast number of materials. However, for many such materials, there is no smoking gun evidence of altermagnetic order, and many predictions were subsequently falsified \cite{liu2024absence, mazinInverseLieb}. Piezomagnetism in the absence of a net dipole moment has recently emerged as an experimental signature of altermagnetism in most altermagnets (the so-called $d$-wave systems) \cite{gati2026PRL, buiarelli_noncollinear_2025}. Piezomagnetism is the magnetic analog of piezoelectricity and refers to a linear relationship between stress (or strain) and net magnetization \cite{dzialoshinskii1958piezomagnetism, moriya1959piezomagnetism, borovik1960piezomagnetism}. It is now well established that for piezomagnetism to occur in the absence of a net dipole moment, a material needs to host ferroically ordered higher-rank multipolar order parameters (such as magnetic octupoles), which render them altermagnetic \cite{bhowal_ferroically_2024, buiarelli_noncollinear_2025, fernandes_topological_2024}. 

Altermagnetism was originally proposed for collinear systems and verified by first-principles calculations that did not consider spin-orbit coupling (SOC) \cite{Hayami2019Nov, yuan2020giant, mazin2021prediction}. Although neglecting SOC complicates precise quantitative predictions of physical observables, this approach clearly isolates nonrelativistic properties—those independent of the SOC energy scale. As a result, spin groups (as opposed to magnetic groups) were rediscovered and tabulated as a means to predict the symmetry of nonrelativistic spin splittings and the resultant macroscopic observables, such as piezomagnetism ~\cite{Smejkal2022Beyond, Chen2024enumeration}. Recent work on spin groups indicates that noncollinear magnetic order can also give rise to altermagnetism, and thus piezomagnetism, even in the absence of SOC \cite{etxebarria2025crystal, Liu2026symmetry}. In such \textit{noncollinear altermagnets}, the only role of SOC is to pick a global coordinate system for the magnetic order with respect to the crystallographic axes -- which it does in collinear altermagnets as well. For a given magnetic order, the qualitative role of SOC in the macroscopic response is therefore independent of the order being collinear or not. Additionally, while SOC is often considered to be necessary to stabilize noncollinear magnetism, geometric frustration is in fact responsible for many well-known noncollinear ordered states, such as the $\sqrt{3}\times \sqrt{3}$ ground state of kagome antiferromagnets \cite{hanely2009} and the all-in-all-out phase on the pyrochlore lattice \cite{Shinaoka2105}. A system with negligible SOC may therefore be a (strongly) noncollinear altermagnet and host nonrelativistic spin splitting, as well as piezomagnetism. 

\begin{figure*}[]
\begin{center}
\includegraphics[scale=0.73]{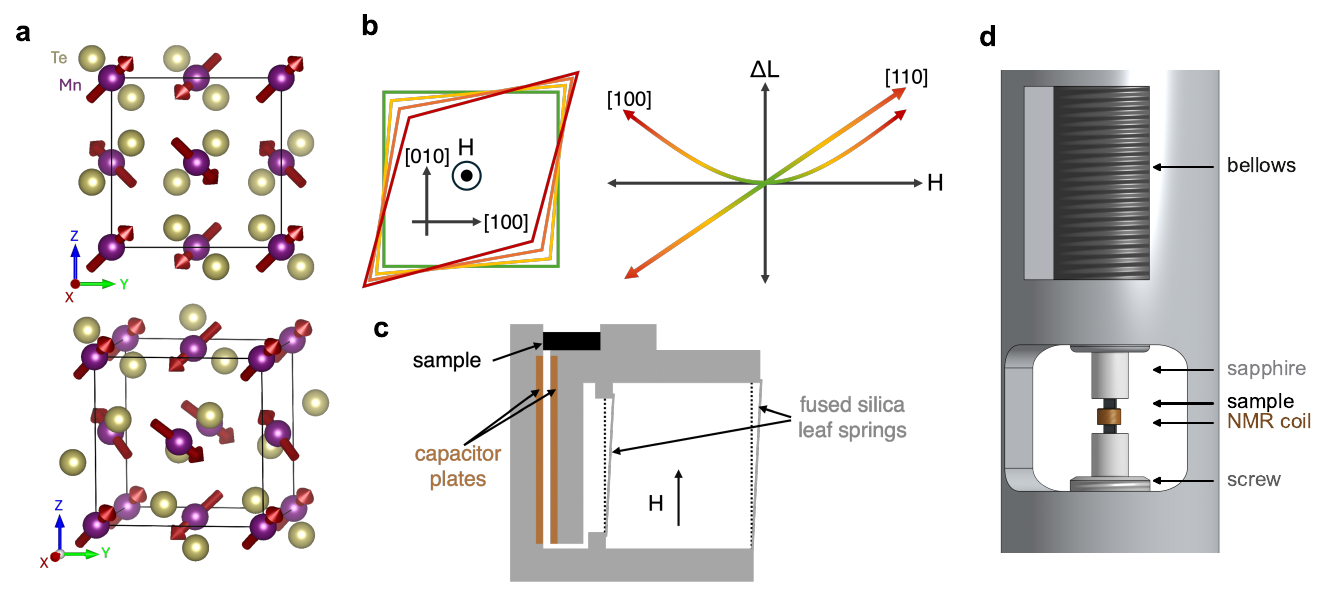}
\caption{ \textbf{Magnetic structure and experimental setup.} \textbf{a}, Schematic crystal and noncollinear magnetic spin structures of MnTe$_2$. \textbf{b}, Sketch of linear magnetostriction in a cubic sample. An applied out-of-plane magnetic field ({\bf H} $||$ [001]) induces a shear strain resulting in elongation/contraction along \lbrack110\rbrack. \textbf{c}, Schematic of dilatometry cell, which was made out of fused silica, with copper sputtered on two surfaces to form the capacitor plates. The magnetic field was applied vertically. \textbf{d}, The NMR strain cell setup uses pressurized helium gas to apply unidirectional stress on the sample placed between two sapphire anvils. The signal is captured by the NMR coil.   \label{Cell_Schematic}}

\end{center}
\end{figure*}

Here we identify the transition-metal ditelluride MnTe$_2$ as an example of such a system. MnTe$_2$ has the so-called pyrite structure, where the four Mn ions in the primitive cell occupy the corner and face-center positions. In the presence of nearest-neighbor antiferromagnetic interactions, such a face-centered lattice of Mn ions is a classic example of geometric frustration \cite{henley1987, anderson_generalizations_1950}. Below $T_N$ = 87~K, the Mn$^{2+} (S = 5/2)$ moments point along \{111\} directions of the cubic unit cell, resulting in a perfectly compensated non-coplanar magnetic state, as shown in Fig. \ref{Cell_Schematic}\textbf{a} \cite{MnTe2NeutronPRB1997}. This magnetic order breaks \textit{PT} symmetry, resulting in an altermagnetic state that displays plaid-like spin splitting of electronic bands and chirality splitting of magnon bands \cite{zhu2024MnTe2plaid, wulferding2026plaid}. 
%
%
We quantify both the direct and inverse piezomagnetic effects using complementary experimental probes and show that the responses are consistent with symmetry analysis. Furthermore, we use first-principles calculations to obtain microscopic insight: we demonstrate that MnTe$_2$ exhibits spin-polarized electronic bands, consistent with expectations for an altermagnetic state, and in agreement with recent experimental work \cite{zhu2024MnTe2plaid}. Significantly, our calculations provide a quantitative description of the experimental results, and they show that shear strain leads to a considerable tilting of spins and to a large bulk ferromagnetic moment even in the absence of SOC.

\subsection*{Theoretical Background}

Piezomagnetism is the linear relation between a net uncompensated magnetic moment and applied stress $\boldsymbol{\sigma}$. Conversely, inverse piezomagnetism (or linear magnetostriction) is the linear coupling between an applied magnetic field $\mathbf{H}$ and the resultant strain  $\boldsymbol{\epsilon} $,
\begin{equation}
    \epsilon_{jk} = \Lambda_{ijk} H_i,
\end{equation}
where $\boldsymbol{\epsilon}$ is written as a six-component vector in Voigt notation. Notice that the direct and inverse effects are closely related through the isothermal Maxwell relation  $(d\epsilon_{ij}/dH_k)_{T,\mathbf{\sigma}}=\mu_0/V (dM_k/d\sigma_{ij})_{T,\mathbf{H}}$, where $V$ is the unit cell volume and  $\mu_0$ the permeability of free space.
%
%
The magnetic order parameter of MnTe$_2$ transforms as the irreducible representation $A_g^-$ ($m\Gamma_1^+$), which leads to the magnetic space group $Pa\bar{3}.1$ (\#205.33)\cite{MnTe2NeutronPRB1997, gallego2016magndata}. This makes MnTe$_2$ an extremely rare example of a cubic altermagnet, despite the highly noncollinear magnetic order. 
This symmetry constrains the inverse piezomagnetic tensor $\boldsymbol{\Lambda}$ to a single independent component $\lambda$, such that a magnetic field applied along any cubic $\{100\}$ axis induces a shear strain in the plane perpendicular to it.

We perform measurements of both magnetostriction and piezomagnetism using high-sensitivity dilatometry and nuclear magnetic resonance, respectively. In the magnetostriction experiments, we focus on the response to a field $H_z$ applied along [001], which induces a small shear strain $\epsilon_{xy}$. Since $\epsilon_{xy}$ transforms as a component of $T_g^+$ ($\Gamma_4^+$) and $H_z$ as a component of $T_g^-$ ($m\Gamma_4^+$), the Landau free energy up to second order in the small parameters takes the form
\begin{equation}
    \mathcal{F} = \frac{\alpha_N}{2}N^2 + \frac{\alpha_\epsilon}{2}\epsilon^2_{xy} + \frac{\alpha_H}{2}H_z^2 + \gamma N\epsilon_{xy}H_z,
\end{equation}
where $N$ is the altermagnetic order parameter and $\alpha_i$ are expansion coefficients for the $i=N$ (magnetic order), $H$ (magnetic susceptibility) and $\epsilon$ (elastic) terms. The saddle-point condition gives a linear relation between the strain and the external magnetic field,
\begin{equation}
    \epsilon_{xy} = -\frac{\gamma}{\alpha_\epsilon}\, N H_z.
\end{equation}
The inverse piezomagnetic coupling is therefore proportional to $N$ and should inherit its temperature dependence. In what follows, we denote its zero-temperature value $\lambda_0$. Note that a third-order term in the free energy that is proportional to $\epsilon_{xy}H_z^2$ is always allowed by symmetry and leads to a magnetostriction contribution that is quadratic in the magnetic field at all temperatures.

\begin{figure*}[]
\begin{center}
\includegraphics[scale=0.74]{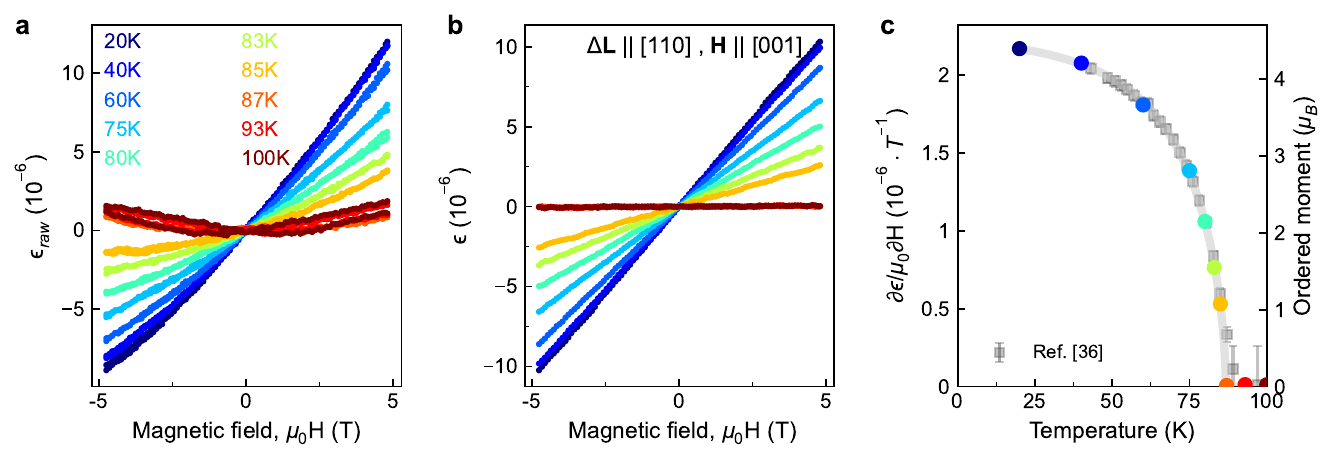}
\caption{\textbf{Transverse magnetostriction measurements for magnetic field along [100] and $\Delta L$ along [110].} \textbf{a}, Magnetic field dependence of the relative deformation. The response is predominantly linear deep in the altermagnetic state, with a small quadratic contribution. \textbf{b}, The antisymmetric component of the relative deformation. The response is linear with respect to the magnetic field. \textbf{c}, Temperature dependence of the slope of the antisymmetric component (left axis) compared with the magnitude of the ordered moment from neutron diffraction (right axis, light gray)\cite{MnTe2NeutronPRB1997}. The gray line is a guide to the eye. The colors in (b,c) correspond to those in (a). \label{110} }

\end{center}
\end{figure*}

\subsection*{Dilatometry Results} 

Figure \ref{Cell_Schematic}\textbf{b} shows a sketch of the lattice response of cubic MnTe$_2$ to a magnetic field along [001]. For this geometry, an in-plane shear strain is induced with a magnitude that is proportional to the applied field. In other words, there is a linear magnetostriction component along [110], whereas magnetostriction is purely quadratic along [100]. As shown in Fig. \ref{110}\textbf{a} and Extended Data Fig. \ref{100}, our data confirm these predictions: along [110], the magnetostriction is dominated by a linear magnetic field dependence, while along [100] the response is purely quadratic. A small quadratic contribution is also observed along [110]; this is expected in any material regardless of symmetry and carries no information about the magnetic state, unlike the linear term, which is the genuine signature of altermagnetism. This component is removed by antisymmetrization of the data,  $\frac{\epsilon(H)-\epsilon (-H)}{2} $, which results in robust linear behavior, $\epsilon(H) = \lambda H$ (Fig. \ref{110}\textbf{b}). Moreover, the inverse piezomagnetic coupling coefficient $\lambda$ follows a temperature dependence reminiscent of an order parameter, with a critical exponent in the 0.31-0.38 range, depending on the fit range (Fig. \ref{110}\textbf{c}), consistent with the expectation for a 3D magnet \cite{chaikin1995principles}. 
In fact, as shown in Fig. \ref{110}\textbf{c}, our result is in excellent agreement with a prior direct neutron diffraction measurement of the temperature dependence of the ordered moment \cite{MnTe2NeutronPRB1997}. 
By extrapolating to zero temperature, we obtain the saturation value $\lambda_{0,\rm{dil}}=2.274(5)\times10^{-6}\textrm{ T}^{-1}$. 



%
The low-temperature magnetostriction data exhibit a linear (i.e., piezomagnetic) response at least up to 16 T (Extended Data Fig. \ref{16Train}). This linear range is much larger than in altermagnets such as CoF$_2$ \cite{borovik1963linear}, and DyFeO$_3$ \cite{zvezdin1985linearDyFeO3}, in which magnetic domains switch at rather small fields. 
It is likely that domain switching, if possible at all in MnTe$_2$, occurs only at very high fields, as observed, \textit{e.g.}, in UO$_2$ \cite{jaime2017UO2}. Additionally, we did not observe any domain switching at $T = 85$ K, just below the magnetic transition, for magnetic fields up to 16 T. This is an indication of a weak coupling between the domains and the applied magnetic field. We note, however, that the linear magnetostriction is weak close to the transition because it scales with the magnetic order parameter, which makes domain switching more difficult to detect. However, it is possible to manipulate the domain population by cooling through $T_N$ in a training field. When cooling in a nominally zero field, the linear magnetostriction generally has a positive slope, likely due to a small negative remanent field of our superconducting magnet; yet upon cooling from 90 K in a magnetic field of 500 G, we obtain a magnetic state exhibiting linear magnetostriction with a negative slope (Extended Data Fig. \ref{16Train}). Additionally, we observe small changes in the slope in different positive and negative field cooling runs (Extended Data Fig. \ref{16Train}). This indicates the presence of multiple magnetic domains and that the intrinsic inverse piezomagnetic coupling coefficient for a single domain is likely larger than the value $\lambda_{0,\rm{dil}}=2.274(5)\times10^{-6}\textrm{ T}^{-1}$ extracted from our dilatometry measurements.  


\begin{figure*}[]
\begin{center}
\includegraphics[scale=0.5]{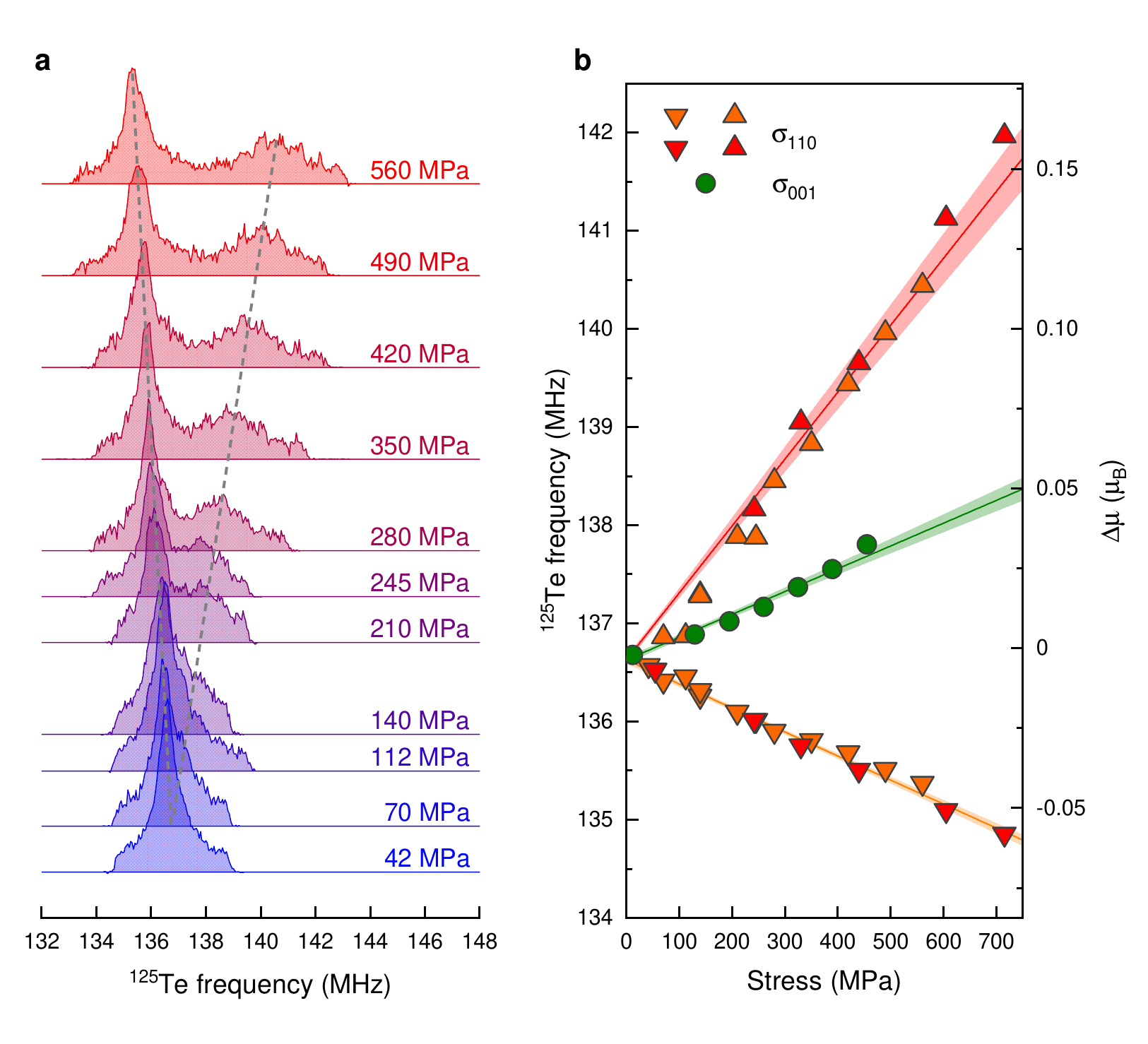}
\caption{\textbf{Nuclear magnetic resonance measurements of the stress-induced ferromagnetic moment.} \textbf{a}, Magnetic resonance spectra of the $^{125}$Te nucleus and its dependence on stress along the [110] direction, measured at 14 K. A splitting of the line is observed, which demonstrates the presence of a ferromagnetic moment along [001]. \textbf{b}, Stress dependence of the resonance frequencies. Triangles correspond to stress along [110], with measurements for two samples superimposed (red and orange symbols). Circles correspond to stress along [001], which does not cause the line to split. The effective moment change (right axis) is obtained using a hyperfine coupling of 32.5~MHz/$\rm \mu_B$. \label{NMR} }

\end{center}
\end{figure*}

\begin{table*}[t]
\centering
\begin{tabular}{@{}l S[table-format=2.2(1)] S[table-format=1.4(1)] S[table-format=1.3(1)]@{}}
\toprule
 & {$\partial M_z/\partial \epsilon_{xy}$} & {$\partial M_z/\partial \sigma_{xy}$} & {$\lambda_0=\partial \epsilon_{xy}/\partial H_z$} \\
 & {($\mu_B$\,u.c.$^{-1}$)} & {($\mu_B$\,GPa$^{-1}$\,u.c.$^{-1}$)} & {($10^{-6}$\,T$^{-1}$)} \\
\midrule
Dilatometry                 & 6.49(2) & 0.0821(2) & \bfseries 2.274(5) \\
Nuclear Magnetic Resonance  & 11.1(8) & \bfseries 0.14(1)   & 3.9(3)   \\
Density Functional Theory   & \bfseries 14.7    & 0.19      & 5.1      \\
\bottomrule
\end{tabular}
\vspace{6pt}
\caption{\textbf{Comparison of direct and inverse piezomagnetic coefficients.}
Comparison of dilatometry, NMR, and DFT results for the (extrapolated) zero-temperature piezomagnetic response of MnTe$_2$. Bold values are directly measured or calculated, while the
rest are derived via the Maxwell relation and elastic constant $C_{44}$.}
\label{tab:comparison}
\end{table*}

\subsection*{NMR Results, Comparison with Dilatometry and Theory}

To independently probe the direct piezomagnetic effect and circumvent the domain sensitivity, we determine the stress-induced moment using nuclear magnetic resonance (NMR), a bulk local probe insensitive to mesoscopic domain formation. Without any applied stress, the internal fields due to the ordered Mn moments yield a single $^{125}$Te resonance around 136.6~MHz (Fig. \ref{NMR}\textbf{a}). Since the ordered moment in MnTe$_2$ is known to be $\rm 4.2~\mu_B$~\cite{MnTe2NeutronPRB1997}, the effective hyperfine coupling for $^{125}$Te is therefore 32.5~MHz/$\rm \mu_B $, where $\rm \mu_B$ is the Bohr magneton. Uniaxial stress applied along [110] splits the resonance into two components with approximately equal weights, and the splitting is proportional to stress (Fig. \ref{NMR}\textbf{a,b}). In contrast, stress along [100] does not split the line, but induces a simple linear shift (Fig. \ref{NMR}\textbf{b} and Extended Data Fig. \ref{MoreNMR}\textbf{a}). This shift most likely originates from the fact that stress along any direction also induces a homogeneous compression component, which enhances orbital overlap and increases the ordered moment. The splitting for [110] stress is fully consistent with the appearance of an induced net moment along [001], and we estimate its size as $0.14(1)$ $\rm \mu_B$/GPa using the effective hyperfine coupling. Importantly, NMR is a local probe, which implies that  the NMR line splitting is independent of the sign of the moment and, therefore, insensitive to domain effects. 
We can employ the Maxwell relation to predict the stress-induced piezomagnetic moment from the dilatometry-derived $\lambda_{0,\rm{dil}}$ and compare with the NMR result. The calculation gives $\partial M_z/\partial \sigma_{xy} = 0.0821(2)~\rm \mu_B/$GPa, which is in good agreement with the value obtained from NMR (Table \ref{tab:comparison}), given that the presence of domains is expected to decrease the macroscopic magnetostriction.

\begin{figure}
    \centering
    \includegraphics[width=0.99\linewidth]{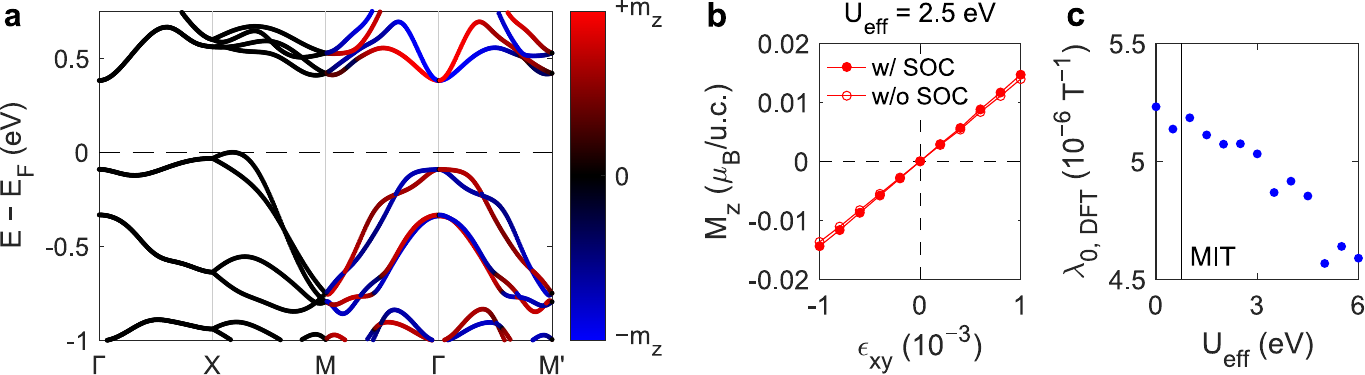}
    \caption{\textbf{First-principles calculations.} \textbf{a}, Electronic band structure of MnTe$_2$, colored by the spin magnetization $m_z$. The sign of $m_z$ reverses between the $\Gamma$--M and $\Gamma$--M' directions, the hallmark of altermagnetic $d$-wave spin splitting. \textbf{b}, Strain-induced magnetic moment per unit cell, $M_z$, vs. shear strain, $\epsilon_{xy}$, computed for $U_{\rm eff} = 2.5$~eV. The linear dependence confirms the symmetry-allowed piezomagnetic coupling even without spin-orbit coupling (w/o SOC). \textbf{c}, Inverse piezomagnetic coupling coefficient $\lambda_{0,DFT}$ extracted from DFT vs. $U_{\rm eff}$. The metal-insulator transition (MIT) is marked by the black vertical line.}
    \label{fig:dft}
\end{figure}

In order to elucidate the origin of the stress-induced ferromagnetic moment, we perform first-principles calculations based on density functional theory (DFT). The results confirm altermagnetic $d$-wave spin splitting of the electronic bands, visible in Fig.~\ref{fig:dft}\textbf{a} as the characteristic sign reversal of spin magnetization $\langle m_z\rangle$ of individual bands between the $\Gamma$--M (1/2, 1/2, 0) and $\Gamma$--M' (1/2, -1/2, 0) directions. Self-consistent calculations at different values of fixed strain show that the net dipole moment is linear in strain, with a slope of $\partial M_z/\partial \epsilon_{xy} \approx 14.7~\mu_B$ per unit cell (Fig. \ref{fig:dft}\textbf{b}). Importantly, this slope does not change significantly when SOC is excluded, underlining that the piezomagnetism in this magnetic space group is nonrelativistic. Since the shear strain is connected to stress via $\sigma_{xy} = C_{44}\epsilon_{xy}$, with $C_{44}$ the shear modulus, it is straightforward to compare this value with our experimental results. 
%
%
Using the DFT calculated value of $C_{44} \approx 79$~GPa, we obtain a stress-induced moment of 0.19~$\rm \mu_B$/GPa. The agreement with the NMR value is excellent (Table \ref{tab:comparison}), especially since DFT is a static mean-field theory and does not capture the quantum and thermal fluctuations that suppress the canted moment in frustrated magnetic systems. The inverse piezomagnetic coupling coefficient is again obtained from the Maxwell relation, with the result $\lambda_{0,\rm{DFT}} \approx 5.1\times 10^{-6}$~T$^{-1}$. As shown in Fig.~\ref{fig:dft}\textbf{c}, varying the effective Hubbard interaction $U_{\rm eff}$ does not significantly affect the results, although an increase in $U$ brings the values even closer to experiment. In addition to the inaccuracy in the exchange-correlation energy (and hence the Hubbard $U$), thermal fluctuations as well as possible extrinsic factors related to the sample quality may be responsible for the disagreement between DFT and NMR results. Importantly, the calculations correctly reproduce the symmetry-allowed nature of the piezomagnetic response, lending theoretical support to the interpretation of the experimental results.

\subsection*{Implications of the Present Work}

Our work has several important implications, both for the fundamental understanding of altermagnetism as well as for potential applications. First, the NMR measurements of the piezomagnetic effect represent, to our knowledge, the first microscopic determination of piezomagnetism using a bulk local probe, thereby opening new avenues for probing altermagnetism in other material systems. Second, the simultaneous observation of both direct and inverse linear piezomagnetic effects, together with the characteristic temperature dependence of the magnetostriction coupling, positions cubic MnTe$_2$ as a model altermagnet that can be quantitatively understood through first-principles modeling. Third, the observed strain-induced moment is significantly larger than in prototypical altermagnets such as $\text{MnTe}$ \cite{PRM_MnTe_piezomagnetism} or candidate systems like $\text{CoF}_2$ and $\text{MnF}_2$ \cite{gati2026PRL}, and comparable in magnitude to materials with strong spin frustration, such as $\text{Mn}_3\text{Ir}$ \cite{Mn3IrGiant} and $\text{Mn}_3\text{NiN}$ \cite{giantMn3NiN}. The high strain sensitivity despite a fully compensated zero-strain magnetic state establishes of MnTe$_2$ as a promising platform to explore strain-controlled piezomagnetic switching of magnetic states \cite{bao2026piezomagnetic}. The finding of a large strain-induced moment also demonstrates that combining shear strain with noncollinear spin configurations can substantially boost the piezomagnetic response. Beyond its fundamental significance, this insight is promising for both static and dynamic sensing (e.g., strong magnetic moments induced by surface acoustic waves) and for altermagnetic domain manipulation via coupled strain and magnetic fields. Finally, given that $\text{MnTe}_2$ is a model narrow-gap semiconductor altermagnet \cite{kasai1981thermal} with exceptional strain sensitivity, this structurally simple system provides an ideal platform for the exploration of strain-induced electronic transport phenomena in spintronic devices, such as the recently proposed elasto-Hall effect~\cite{elastohallconductvity}.



\section*{Methods}

\subsection*{Samples} Single crystals of MnTe$_2$ were synthesized by vapor transport using iodine as the transport agent. Mn (99.99\%) and Te (99.999\%) powders along with I$_2$ pieces were sealed in a quartz ampoule, which was maintained at 600 $^\circ$C for 10 days with a temperature gradient of 60 $^\circ$C over 18 cm using two zones of a three-zone furnace. We obtained shiny, plate-like crystals with linear dimensions of 3-5 mm and clear facets. We performed a room-temperature powder x-ray diffraction (XRD) measurement of a pulverized MnTe$_2$ crystal and carried out Rietveld refinement of the data [Extended Data Fig. \ref{SElaue}\textbf{a}] to confirm that the crystals grow in a cubic structure with space group Pa$\bar{3}$, with lattice parameters \textit{a = b = c = }6.949(4) \AA.  Samples were aligned to the desired orientations using a Laue diffractometer [Extended Data Fig. \ref{SElaue}\textbf{c}], with an angular uncertainty of approximately $3^\circ$. Magnetic characterization was performed with a Quantum Design, Inc. MPMS instrument [Extended Data Fig. \ref{SElaue}\textbf{b} ].

\subsection*{Dilatometry} We built a fused silica capacitive dilatometer based on the design reported in ref. \cite{neumeier2008capacitive}  to measure thermal expansion and magnetostriction (see Fig. \ref{Cell_Schematic}\textbf{b}). The capacitance was measured with an IET/QuadTech 7600 Plus precision LCR meter using a 1 V and 12.345 kHz excitation signal. Figure \ref{Cell_Schematic}\textbf{c} shows the geometry for a sample mounted in the dilatometer cell.  

The most general expression for the fractional change in length of a sample measured with our cell is
\newline
\begin{equation}
\label{general_DL_over_L}
        \epsilon\equiv\frac{\Delta L}{L}=\frac{A\epsilon_0}{L_{\textrm{ref}}}\left[\frac{1}{C(T,H)}- \frac{1}{C_{\textrm{ref}}} \left(1 + \delta L_{\textrm{fs}}\right)\right]\\+ \delta L_{\textrm{fs}},
\end{equation}
\newline
where $A$ is the effective plate area, $\epsilon_0$ is the permittivity of free space, $L_{\textrm{ref}}=L(T_{\textrm{ref}},H=0)$ is the length of the sample at the reference temperature $T_{\textrm{ref}}$ and zero applied magnetic field, $C(T,H)$ is the sample contribution to the capacitance at temperature $T$ and applied magnetic field $H$, $C_{\textrm{ref}}=C(T_{\textrm{ref}},H=0)$, and $\delta L_{\textrm{fs}}=\frac{\Delta L_{\textrm{fs}}(T,H)}{L_{\textrm{fs}}(T,H)}$ is the thermal expansion of fused silica obtained from ref. \cite{neumeier2008capacitive}. The thermal expansion in zero applied magnetic field, $\epsilon(T)$, can be easily obtained from Eqn. \ref{general_DL_over_L}. We used $T_{\textrm{ref}}=293 \textrm{ K}$ for our calculations. The magnetostriction at fixed temperature, $\epsilon(H)$, was also obtained from Eqn. \ref{general_DL_over_L} and has the simple form
\begin{equation}
\epsilon(H) = \frac{A\epsilon_0}{L_{\textrm{ref}}}\left[\frac{1}{C(T_{\textrm{ref}},H)} - \frac{1}{C(T_{\textrm{ref}},0)}\right],
\end{equation}
\newline 
since $\delta L_{\textrm{fs}}$ has no magnetic field dependence.
The raw measured capacitance for a sample contains contributions from the dilatometer itself, which need to be subtracted. The dilatometer's contribution to the capacitance was determined by installing a piece of fused silica and measuring the temperature dependence of the resulting capacitance, $C_{\textrm{fs}}(T)$. Using this value, the sample capacitance could be extracted from raw capacitance data using the expression
\begin{equation}
    C(T) = C_{\textrm{raw}}(T)-C_{\textrm{raw}}(T_{\textrm{ref}})\times\left[\frac{C_\textrm{fs}(T)-C_\textrm{fs}(T_{\textrm{ref}})}{C_{\textrm{fs}}(T_{\textrm{ref}})}\right].
\end{equation}
The effective plate area was determined by measuring a high-purity copper sample and using Eqn. \ref{general_DL_over_L} to compare to published thermal expansion data for copper \cite{Kroeger1977}.

The thermal expansion coefficient, $\alpha(T)$, was calculated by taking the temperature derivative of the thermal expansion:
\begin{equation}
    \alpha(T)=\frac{d\epsilon}{dT}.
\end{equation}
\newline
The temperature dependencies of the relative thermal expansion and thermal expansion coefficient for a [100]-oriented MnTe$_2$ sample shows a magnetism-induced contraction at $T_N$ [Extended Data Fig. \ref{SElaue}\textbf{d}]. 

The magnetostriction measurements were performed at various temperatures and slope of the strain versus magnetic field was used to determine the inverse piezomagnetic coupling coefficient, $\lambda_0$.  We used the Maxwell relation,
\begin{equation}
\lambda_0 = \frac{1}{\mu_0} \left( \frac{\partial \varepsilon_{xy}}{\partial H_z} \right)_{T,\sigma_{xy}} = \frac{1}{a^3C_{44}} \left(\frac{\partial M_z}{\partial \varepsilon_{xy}}\right)_{T,H_z},
\end{equation}
to extract the stressed induced magnetic moment to compare with DFT and NMR measurements. Here $a \approx 6.95$ \AA~is the lattice parameter of MnTe$_2$.

\subsection*{Nuclear Magnetic Resonance}
We performed pulsed NMR experiments on the $^{125}$Te nucleus in zero external magnetic field using Tecmag spectrometers and a liquid helium cryostat with a variable-temperature insert. A standard Hahn echo pulse sequence was used for all measurements, with sample masses in the mg range. A pneumatic uniaxial strain cell with a maximum force on the sample of 350~N was used, similar to previous work~\cite{AN2022PRL
}. The stress dependence of the spectra was determined at 14 K, where the spin-lattice relaxation time $T_1$ is on the order of 10 ms; due to the extremely fast increase of $T_1$ upon cooling (Extended Data Fig. \ref{MoreNMR}\textbf{b}), measurements at lower temperatures are difficult. Conversely, the spin-spin relaxation time decreases sharply above about 15 K (Extended Data Fig. \ref{MoreNMR}\textbf{c}), which again suppresses the signal. Temperatures between 10 and 20 K are therefore most favorable.

\subsection*{Density Functional Theory} We simulated the noncollinear magnetic phase of MnTe$_2$ using Density Functional Theory (DFT) as implemented in Abinit v9.10.3, within the local density approximation (LDA) and using plane-augmented waves (PAW)~\cite{torrent2008, gonze2020} with Jollet-Torrent-Holzwarth pseudo-potentials~\cite{Jollet2014Generation}. We used a $\Gamma$-centered 12$\times$12$\times$12 \textit{k}-point grid and energy cutoffs of 550 eV for the plane-wave basis and 1100 eV for the spherical grid inside the atomic augmentation spheres. In order to obtain an insulating state, we used DFT+$U$ in the fully localized limit~\cite{amadon2008}, trying different values of $U$ and fixing $J=0.5$ eV. In the investigation of strain-induced magnetism, we applied various magnitudes of symmetry-breaking strain ($\epsilon_{xy}$) and computed the resultant net magnetic moment per unit cell by performing self-consistent calculations. While atoms are expected to be displaced from their high-symmetry Wyckoff positions under strain, we did not take into account this ion-mediated contribution.

\section*{Acknowledgments}
 We thank Xianghan Xu, Chris Leighton, and Marin Lukas for comments and discussions. The work at the University of Minnesota was funded by the US Department of Energy through the University of Minnesota Center for Quantum Materials, under Grant No. DE-SC-0016371. The work at the University of Zagreb was funded by the Croatian Science Foundation under grant no. IP-2026-6146.

\section*{Author Contributions}
S.S., D.P., T.B., and M.G. designed the research. S.S. grew the single crystals. S.S. and R.S. performed the dilatometry experiments. I.J. and D.P. performed the NMR experiments.  L.B. and T.B. carried out the theoretical calculations. S.S., R.S., I.J., and D.P. analyzed the data. All authors wrote the paper.

\bibliography{References}

@article{neumeier2008capacitive,
  title={Capacitive-based dilatometer cell constructed of fused quartz for measuring the thermal expansion of solids},
  author={Neumeier, JJ and Bollinger, RK and Timmins, GE and Lane, CR and Krogstad, RD and Macaluso, J},
  journal = {Review of Scientific Instruments},
  volume = {79},
  number = {3},
  pages = {033903},
  year = {2008},
  doi = {10.1063/1.2884193},
  publisher = {AIP Publishing}
}

@article{SmejkalPRX2022,
  title = {Emerging Research Landscape of Altermagnetism},
  author = {\ifmmode \check{S}\else \v{S}\fi{}mejkal, Libor and Sinova, Jairo and Jungwirth, Tomas},
  journal = {Phys. Rev. X},
  volume = {12},
  issue = {4},
  pages = {040501},
  numpages = {27},
  year = {2022},
  month = {Dec},
  publisher = {American Physical Society},
  doi = {10.1103/PhysRevX.12.040501},
}

@article{Kroeger1977,
  title = {Absolute linear thermal-expansion measurements on copper and aluminum from 5 to 320 K},
  author = {Kroeger, F. R. and Swenson, C. A.},
  journal = {J. Appl. Phys.},
  volume = {48},
  pages = {853-864},
  year = {1977},
  doi = {10.1063/1.323746},
}

@article{kasai1981thermal,
  title={Thermal Expansion of \textrm{MnTe$_2$}},
  author={Kasai, Naoko and Waki, Shinya and Ogawa, Shinji},
  journal={Journal of the Physical Society of Japan},
  volume={50},
  number={10},
  pages={3303--3307},
  year={1981},
  doi={10.1143/JPSJ.50.3303},
  publisher={The Physical Society of Japan}
}

@article{MnTe2NeutronPRB1997,
  title = {Noncollinear magnetic structure of \textrm{MnTe$_2$}},
  author = {Burlet, P. and Ressouche, E. and Malaman, B. and Welter, R. and Sanchez, J. P. and Vulliet, P.},
  journal = {Phys. Rev. B},
  volume = {56},
  issue = {21},
  pages = {14013--14018},
  numpages = {0},
  year = {1997},
  month = {Dec},
  publisher = {American Physical Society},
  doi = {10.1103/PhysRevB.56.14013},
}

@article{krempasky2024altermagnetic,
  title={Altermagnetic lifting of {Kramers} spin degeneracy},
  author={Krempask{\`y}, Juraj and {\v{S}}mejkal, L and D’souza, SW and Hajlaoui, M and Springholz, G and Uhl{\'\i}{\v{r}}ov{\'a}, K and Alarab, F and Constantinou, PC and Strocov, V and Usanov, D and others},
  journal={Nature},
  volume={626},
  number={7999},
  pages={517--522},
  year={2024},
  doi={10.1038/s41586-023-06907-7},
  publisher={Nature Publishing Group UK London}
}

@article{PRM_MnTe_piezomagnetism,
  title = {Piezomagnetic properties in altermagnetic {MnTe}},
  author = {Aoyama, Takuya and Ohgushi, Kenya},
  journal = {Phys. Rev. Mater.},
  volume = {8},
  issue = {4},
  pages = {L041402},
  numpages = {5},
  year = {2024},
  month = {Apr},
  publisher = {American Physical Society},
  doi = {10.1103/PhysRevMaterials.8.L041402},
}

@article{zhu2024MnTe2plaid,
  title={Observation of plaid-like spin splitting in a noncoplanar antiferromagnet},
  author={Zhu, Yu-Peng and Chen, Xiaobing and Liu, Xiang-Rui and Liu, Yuntian and Liu, Pengfei and Zha, Heming and Qu, Gexing and Hong, Caiyun and Li, Jiayu and Jiang, Zhicheng and others},
  journal={Nature},
  volume={626},
  number={7999},
  pages={523--528},
  year={2024},
  doi={10.1038/s41586-024-07023-w},
  publisher={Nature Publishing Group UK London}
}

@article{LiuPRLChiralSpilting,
  title = {Chiral Split Magnon in Altermagnetic {MnTe}},
  author = {Liu, Zheyuan and Ozeki, Makoto and Asai, Shinichiro and Itoh, Shinichi and Masuda, Takatsugu},
  journal = {Phys. Rev. Lett.},
  volume = {133},
  issue = {15},
  pages = {156702},
  numpages = {6},
  year = {2024},
  month = {Oct},
  publisher = {American Physical Society},
  doi = {10.1103/PhysRevLett.133.156702},
}

@article{fualtermagneticspintronics,
  title={All-electrically controlled spintronics in altermagnetic heterostructures},
  author={Fu, Pei-Hao and Lv, Qianqian and Xu, Yong and Cayao, Jorge and Liu, Jun-Feng and Yu, Xiang-Long},
  journal={npj Quantum Materials},
  year={2025},
  volume={10},
  pages={111},
  doi={10.1038/s41535-025-00827-7},
  publisher={Nature Publishing Group UK London}
}

@article{hu2025spinhallncalm,
  title={Spin {Hall} and {Edelstein} effects in chiral non-collinear altermagnets},
  author={Hu, Mengli and Janson, Oleg and Felser, Claudia and McClarty, Paul and Van den Brink, Jeroen and G. Vergniory, Maia},
  journal={Nature Communications},
  volume={16},
  number={1},
  pages={8529},
  year={2025},
  doi={10.1038/s41467-025-64271-8},
  publisher={Nature Publishing Group UK London}
}

@article{amadon2008,
  title = {$\ensuremath{\gamma}$ and $\ensuremath{\beta}$ cerium: {$\text{LDA}$} $+$ {$\text{U}$} calculations of ground-state parameters},
  author = {Amadon, B. and Jollet, F. and Torrent, M.},
  journal = {Phys. Rev. B},
  volume = {77},
  issue = {15},
  pages = {155104},
  numpages = {10},
  year = {2008},
  month = {Apr},
  publisher = {American Physical Society},
  doi = {10.1103/PhysRevB.77.155104},
}

@article{torrent2008,
title = {Implementation of the projector augmented-wave method in the {ABINIT} code: Application to the study of iron under pressure},
journal = {Computational Materials Science},
volume = {42},
number = {2},
pages = {337-351},
year = {2008},
issn = {0927-0256},
doi = {10.1016/j.commatsci.2007.07.020},
author = {Marc Torrent and François Jollet and François Bottin and Gilles Zérah and Xavier Gonze}
}

@article{Jollet2014Generation,
title = {Generation of Projector Augmented-Wave atomic data: A 71 element validated table in the XML format},
journal = {Computer Physics Communications},
volume = {185},
number = {4},
pages = {1246-1254},
year = {2014},
issn = {0010-4655},
doi = {https://doi.org/10.1016/j.cpc.2013.12.023},
author = {François Jollet and Marc Torrent and Natalie Holzwarth}
}

@article{gonze2020,
title = {The {Abinitproject}: Impact, environment and recent developments},
journal = {Computer Physics Communications},
volume = {248},
pages = {107042},
year = {2020},
issn = {0010-4655},
doi = {https://doi.org/10.1016/j.cpc.2019.107042},
author = {Xavier Gonze and Bernard Amadon and Gabriel Antonius and Frédéric Arnardi and Lucas Baguet and Jean-Michel Beuken and Jordan Bieder and François Bottin and Johann Bouchet and Eric Bousquet and Nils Brouwer and Fabien Bruneval and Guillaume Brunin and Théo Cavignac and Jean-Baptiste Charraud and Wei Chen and Michel Côté and Stefaan Cottenier and Jules Denier and Grégory Geneste and Philippe Ghosez and Matteo Giantomassi and Yannick Gillet and Olivier Gingras and Donald R. Hamann and Geoffroy Hautier and Xu He and Nicole Helbig and Natalie Holzwarth and Yongchao Jia and François Jollet and William Lafargue-Dit-Hauret and Kurt Lejaeghere and Miguel A.L. Marques and Alexandre Martin and Cyril Martins and Henrique P.C. Miranda and Francesco Naccarato and Kristin Persson and Guido Petretto and Valentin Planes and Yann Pouillon and Sergei Prokhorenko and Fabio Ricci and Gian-Marco Rignanese and Aldo H. Romero and Michael Marcus Schmitt and Marc Torrent and Michiel J. {van Setten} and Benoit {Van Troeye} and Matthieu J. Verstraete and Gilles Zérah and Josef W. Zwanziger}
}

@article{jaime2017UO2,
  title={Piezomagnetism and magnetoelastic memory in uranium dioxide},
  author={Jaime, Marcelo and Saul, A and Salamon, M and Zapf, VS and Harrison, Neil and Durakiewicz, Tomasz and Lashley, Jason Charles and Andersson, DA and Stanek, Christopher Richard and Smith, James L and others},
  journal={Nature communications},
  volume={8},
  number={1},
  pages={99},
  year={2017},
  doi={10.1038/s41467-017-00096-4},
  publisher={Nature Publishing Group UK London}
}

@article{zvezdin1985linearDyFeO3,
  title={Linear magnetostriction and the antiferromagnetic domain structure in dysprosium orthoferrite},
  author={Zvezdin, AK and Zorin, I and Kadomtseva, AM and Krynetskii, IB and Moskvin, AS and Mukhin, AA},
  journal={Zhurnal Eksperimental'noi i Teoreticheskoi Fiziki},
  volume={88},
  number={3},
  pages={1098--1102},
  year={1985}
}

@article{borovik1963linear,
  author = {Borovik-Romanov, A. S. and Yavelov, B. E.},
  title  = {Linear magnetostriction in antiferromagnetic \textrm{CoF$_2$}},
  year   = {1963},
  pages  = {81--83},
  journal = {Physics and Techniques of Low Temperatures (Proc. 3rd Regional Conf., Prague)}, 
}

@article{gallego2016magndata,
  title={MAGNDATA: towards a database of magnetic structures. I. The commensurate case},
  author={Gallego, Samuel V and Perez-Mato, J Manuel and Elcoro, Luis and Tasci, Emre S and Hanson, Robert M and Momma, Koichi and Aroyo, Mois I and Madariaga, Gotzon},
  journal={Applied Crystallography},
  volume={49},
  number={5},
  pages={1750--1776},
  year={2016},
  publisher={International Union of Crystallography},
  doi={10.1107/S1600576716012863}
}

@article{elastohallconductvity,
  title = {Elasto-Hall conductivity and the anomalous Hall effect in altermagnets},
  author = {Takahashi, Keigo and Steward, Charles R. W. and Ogata, Masao and Fernandes, Rafael M. and Schmalian, J\"org},
  journal = {Phys. Rev. B},
  volume = {111},
  issue = {18},
  pages = {184408},
  numpages = {18},
  year = {2025},
  month = {May},
  publisher = {American Physical Society},
  doi = {10.1103/PhysRevB.111.184408},
}

@article{wulferding2026plaid,
  title={Plaid-Like Spin Splitting and Chirality of Magnon Bands in Antiferromagnetic \textrm{MnTe$_2$}},
  author={Wulferding, Dirk and An, Daehyeon and Choi, Jiwon and Moon, Dongmin and Choi, Youngsu and Kuppusamy, Sivasakthi and Krishnamoorthi, Sritharan and Sankar, Raman and Han, Myung Joon and Kim, Se Kwon and others},
  journal={Advanced Science},
  pages={e76555},
  year={2026},
  publisher={Wiley Online Library}
}

@article{moriya1959piezomagnetism,
  title={Piezomagnetism in \textrm{CoF$_2$}},
  author={Moriya, T},
  journal={Journal of Physics and Chemistry of Solids},
  volume={11},
  number={1-2},
  pages={73--77},
  year={1959},
  publisher={Elsevier}
}

@article{Mn3IrGiant,
author = {Zhang, Kaiqi and Ma, Zhijie and Sun, Ying and Shi, Kewen and Wang, Yuyan and Sun, Qisong and Li, YongJie and Deng, Sihao and Cui, Jin and Xia, Zhengcai and Zhao, Weisheng and Wang, Cong},
title = {Superior Thermally Stable Giant Piezomagnetism in Spin Frustrated \textrm{Mn$_3$Ir}},
journal = {Advanced Functional Materials},
volume = {36},
number = {21},
pages = {2424472},
doi = {https://doi.org/10.1002/adfm.202424472},
eprint = {https://advanced.onlinelibrary.wiley.com/doi/pdf/10.1002/adfm.202424472},
year = {2026}
}

@article{giantMn3NiN,
    author = {Boldrin, David and Mihai, Andrei P. and Zou, Bin and Zemen, Jan and Thompson, Ryan and Ware, Ecaterina and Neamtu, Bogdan V. and Ghivelder, Luis and Esser, Bryan and McComb, David W. and Petrov, Peter and Cohen, Lesley F.},
    title = {Giant
Piezomagnetism in \textrm{Mn$_3$NiN}},
    journal = {ACS Applied Materials I\& Interfaces},
    volume = {10},
    number = {22},
    pages = {18863-18868},
    year = {2018},
    month = {05},
    issn = {1944-8244},
    doi = {10.1021/acsami.8b03112},
    eprint = {https://pubs.acs.org/aamick/article-pdf/10/22/18863/7150484/am8b03112.pdf},
}

@article{bao2026piezomagnetic,
  title={Piezomagnetic Switching of Nonvolatile Antiferromagnetic States},
  author={Bao, Xilai and Pylypovskyi, Oleksandr V and Yang, Huali and Xie, Yali and Faurie, Damien and Zighem, Fatih and Weber, Sophie F and Wang, Jiabin and Liang, Jiachen and Xu, Hong and others},
  journal={arXiv preprint arXiv:2604.12786},
  year={2026}
}

@article{AN2022PRL,
  title = {Uniaxial Strain Control of Bulk Ferromagnetism in Rare-Earth Titanates},
  author = {Najev, A. and Hameed, S. and Gautreau, D. and Wang, Z. and Joe, J. and Po\ifmmode \check{z}\else \v{z}\fi{}ek, M. and Birol, T. and Fernandes, R. M. and Greven, M. and Pelc, D.},
  journal = {Phys. Rev. Lett.},
  volume = {128},
  issue = {16},
  pages = {167201},
  numpages = {6},
  year = {2022},
  month = {Apr},
  publisher = {American Physical Society},
  doi = {10.1103/PhysRevLett.128.167201}
}

@book{chaikin1995principles,
  title={Principles of condensed matter physics},
  author={Chaikin, Paul M and Lubensky, Tom C and Witten, Thomas A},
  volume={10},
  year={1995},
  publisher={Cambridge university press Cambridge}
}

@article{hanely2009,
  title = {Long-range order in the classical kagome antiferromagnet: Effective Hamiltonian approach},
  author = {Henley, Christopher L.},
  journal = {Phys. Rev. B},
  volume = {80},
  issue = {18},
  pages = {180401(R)},
  numpages = {4},
  year = {2009},
  month = {Nov},
  publisher = {American Physical Society},
  doi = {10.1103/PhysRevB.80.180401}
}

@article{etxebarria2025crystal,
  title={Crystal tensor properties of magnetic materials with and without spin--orbit coupling. Application of spin point groups as approximate symmetries},
  author={Etxebarria, Jesus and Perez-Mato, J Manuel and Tasci, Emre S and Elcoro, Luis},
  journal={Foundations of Crystallography},
  volume={81},
  number={4},
  pages={317--338},
  year={2025},
  publisher={International Union of Crystallography}
}

@article{haule2025,
  title = {High-Throughput Search for Metallic Altermagnets by Embedded Dynamical Mean Field Theory},
  author = {Wan, Xuhao and Mandal, Subhasish and Guo, Yuzheng and Haule, Kristjan},
  journal = {Phys. Rev. Lett.},
  volume = {135},
  issue = {10},
  pages = {106501},
  numpages = {9},
  year = {2025},
  month = {Sep},
  publisher = {American Physical Society},
  doi = {10.1103/k47t-23gp}
}

@article{vandanbrink2023,
  title={Spin-split collinear antiferromagnets: A large-scale ab-initio study},
  author={Guo, Yaqian and Liu, Hui and Janson, Oleg and Fulga, Ion Cosma and van den Brink, Jeroen and Facio, Jorge I},
  journal={Materials Today Physics},
  volume={32},
  pages={100991},
  year={2023},
  publisher={Elsevier}
}

@article{mazinInverseLieb,
  title={Inverse \textrm{Lieb} materials: Altermagnetism and more},
  author={Chang, Po-Hao and Belashchenko, Kirill D and Mazin, Igor I},
  journal={npj Quantum Materials},
  year={2026},
  publisher={Nature Publishing Group UK London}
}

@article{Chen2024enumeration,
  title = {Enumeration and Representation Theory of Spin Space Groups},
  author = {Chen, Xiaobing and Ren, Jun and Zhu, Yanzhou and Yu, Yutong and Zhang, Ao and Liu, Pengfei and Li, Jiayu and Liu, Yuntian and Li, Caiheng and Liu, Qihang},
  journal = {Phys. Rev. X},
  volume = {14},
  issue = {3},
  pages = {031038},
  numpages = {33},
  year = {2024},
  month = {Aug},
  publisher = {American Physical Society},
  doi = {10.1103/PhysRevX.14.031038}
}

@article{Liu2026symmetry,
  title={Symmetry classification of magnetic orders using oriented spin space groups},
  author={Liu, Yuntian and Chen, Xiaobing and Yu, Yutong and Etxebarria, Jes{\'u}s and Perez-Mato, J Manuel and Liu, Qihang},
  journal={Nature},
  volume={652},
  number={8111},
  pages={869--873},
  year={2026},
  publisher={Nature Publishing Group UK London},
  doi={https://doi.org/10.1038/s41586-026-10401-1}
}

@article{Shinaoka2105,
  title = {Phase Diagram of Pyrochlore Iridates: All-in--All-out Magnetic Ordering and Non-Fermi-Liquid Properties},
  author = {Shinaoka, Hiroshi and Hoshino, Shintaro and Troyer, Matthias and Werner, Philipp},
  journal = {Phys. Rev. Lett.},
  volume = {115},
  issue = {15},
  pages = {156401},
  numpages = {5},
  year = {2015},
  month = {Oct},
  publisher = {American Physical Society},
  doi = {10.1103/PhysRevLett.115.156401}
}

@article{dzialoshinskii1958piezomagnetism,
  title={The problem of piezomagnetism},
  author={Dzialoshinskii, Igor Ekhielevich},
  journal={Sov. Phys. JETP},
  volume={6},
  pages={621},
  year={1958}
}

@article{gati2026PRL,
  title = {Probing Multipolar Order in the Candidate Altermagnet \textrm{MnF$_2$} through the Elastocaloric Effect under Strain},
  author = {Ohlendorf, Rahel and Buiarelli, Luca and Noad, Hilary M. L. and Mackenzie, Andrew P. and Fernandes, Rafael M. and Birol, Turan and Schmalian, J\"org and Gati, Elena},
  journal = {Phys. Rev. Lett.},
  volume = {137},
  issue = {5},
  pages = {056702},
  numpages = {9},
  year = {2026},
  month = {Jul},
  publisher = {American Physical Society},
  doi = {10.1103/svrz-315w}
}

@article{park_impact_2026,
	title = {Impact of strong electronic correlations on altermagnets: {The} case of \textrm{NiS$_2$}},
	volume = {10},
	issn = {2475-9953},
	shorttitle = {Impact of strong electronic correlations on altermagnets},
	doi = {10.1103/pgp6-zlh8},
	language = {en},
	number = {5},
	urldate = {2026-08-27},
	journal = {Physical Review Materials},
	author = {Park, Ina and Birol, Turan and Georges, Antoine and Fernandes, Rafael M.},
	month = may,
	year = {2026},
	pages = {054415},
}

@article{fang_quantum_2024,
	title = {Quantum {Geometry} {Induced} {Nonlinear} {Transport} in {Altermagnets}},
	volume = {133},
	issn = {0031-9007, 1079-7114},
	doi = {10.1103/PhysRevLett.133.106701},
	language = {en},
	number = {10},
	urldate = {2026-05-27},
	journal = {Physical Review Letters},
	author = {Fang, Yuan and Cano, Jennifer and Ghorashi, Sayed Ali Akbar},
	month = sep,
	year = {2024},
	pages = {106701},
}

@article{banerjee_altermagnetic_2024,
	title = {Altermagnetic superconducting diode effect},
	volume = {110},
	issn = {2469-9950, 2469-9969},
	doi = {10.1103/PhysRevB.110.024503},
	language = {en},
	number = {2},
	urldate = {2026-08-27},
	journal = {Physical Review B},
	author = {Banerjee, Sayan and Scheurer, Mathias S.},
	month = jul,
	year = {2024},
	pages = {024503},
}

@article{bhowal_ferroically_2024,
	title = {Ferroically {Ordered} {Magnetic} {Octupoles} in \text{d-Wave} Altermagnets},
	volume = {14},
	doi = {10.1103/PhysRevX.14.011019},
	number = {1},
	urldate = {2025-06-11},
	journal = {Physical Review X},
	publisher = {American Physical Society},
	author = {Bhowal, Sayantika and Spaldin, Nicola A.},
	month = feb,
	year = {2024},
	pages = {011019},
}

@article{buiarelli_noncollinear_2025,
	title = {Noncollinear magnetic multipoles in collinear altermagnets},
	volume = {112},
	issn = {2469-9950, 2469-9969},
	doi = {10.1103/kq6x-7jfc},
	language = {en},
	number = {22},
	urldate = {2026-08-27},
	journal = {Physical Review B},
	author = {Buiarelli, Luca and Fernandes, Rafael M. and Birol, Turan},
	month = dec,
	year = {2025},
	pages = {224442},
}

@article{fernandes_topological_2024,
	title = {Topological transition from nodal to nodeless {Zeeman} splitting in altermagnets},
	volume = {109},
	issn = {2469-9950, 2469-9969},
	doi = {10.1103/PhysRevB.109.024404},
	language = {en},
	number = {2},
	urldate = {2026-08-27},
	journal = {Physical Review B},
	author = {Fernandes, Rafael M. and De Carvalho, Vanuildo S. and Birol, Turan and Pereira, Rodrigo G.},
	month = jan,
	year = {2024},
	pages = {024404},
}

@article{henley1987,
    author = {Henley, Christopher L.},
    title = {Ordering by disorder: Ground‐state selection in fcc vector antiferromagnets},
    journal = {Journal of Applied Physics},
    volume = {61},
    number = {8},
    pages = {3962-3964},
    year = {1987},
    month = {04},
    issn = {0021-8979},
    doi = {10.1063/1.338570},
    eprint = {https://pubs.aip.org/aip/jap/article-pdf/61/8/3962/18609897/3962_1_online.pdf},
}

@article{anderson_generalizations_1950,
	title = {Generalizations of the {Weiss} {Molecular} {Field} {Theory} of {Antiferromagnetism}},
	volume = {79},
	copyright = {http://link.aps.org/licenses/aps-default-license},
	issn = {0031-899X},
	doi = {10.1103/PhysRev.79.705},
	language = {en},
	number = {4},
	urldate = {2026-08-27},
	journal = {Physical Review},
	author = {Anderson, P. W.},
	month = {aug},
	year = {1950},
	pages = {705--710},
}

@article{Smejkal2022Beyond,
  title = {Beyond Conventional Ferromagnetism and Antiferromagnetism: A Phase with Nonrelativistic Spin and Crystal Rotation Symmetry},
  author = {\ifmmode \check{S}\else \v{S}\fi{}mejkal, Libor and Sinova, Jairo and Jungwirth, Tomas},
  journal = {Phys. Rev. X},
  volume = {12},
  issue = {3},
  pages = {031042},
  numpages = {16},
  year = {2022},
  month = {Sep},
  publisher = {American Physical Society},
  doi = {10.1103/PhysRevX.12.031042},
}

@article{Yu2025odd,
  title = {Odd-Parity Magnetism Driven by Antiferromagnetic Exchange},
  author = {Yu, Yue and Lyngby, Magnus B. and Shishidou, Tatsuya and Roig, Merc\`e and Kreisel, Andreas and Weinert, Michael and Andersen, Brian M. and Agterberg, Daniel F.},
  journal = {Phys. Rev. Lett.},
  volume = {135},
  issue = {4},
  pages = {046701},
  numpages = {7},
  year = {2025},
  month = {Jul},
  publisher = {American Physical Society},
  doi = {10.1103/zk69-k6b2},
}

@article{borovik1960piezomagnetism,
  title={Piezomagnetism in the antiferromagnetic fluorides of cobalt and manganese},
  author={Borovik-Romanov, AS},
  journal={Sov. Phys. JETP},
  volume={11},
  number={4},
  pages={786},
  year={1960}
}

@article{jungwirth2025altermagnetism,
  title={Altermagnetism: An unconventional spin-ordered phase of matter},
  author={Jungwirth, Tom{\'a}{\v{s}} and Fernandes, Rafael M and Fradkin, Eduardo and MacDonald, Allan H and Sinova, Jairo and {\v{S}}mejkal, Libor},
  journal={Newton},
  volume={1},
  number={6},
  year={2025},
  publisher={Elsevier},
  doi={10.1016/j.newton.2025.100162}
}

@article{fedchenko2024observation,
author = {Olena Fedchenko  and Jan Minár  and Akashdeep Akashdeep  and Sunil Wilfred D’Souza  and Dmitry Vasilyev  and Olena Tkach  and Lukas Odenbreit  and Quynh Nguyen  and Dmytro Kutnyakhov  and Nils Wind  and Lukas Wenthaus  and Markus Scholz  and Kai Rossnagel  and Moritz Hoesch  and Martin Aeschlimann  and Benjamin Stadtmüller  and Mathias Kläui  and Gerd Schönhense  and Tomas Jungwirth  and Anna Birk Hellenes  and Gerhard Jakob  and Libor Šmejkal  and Jairo Sinova  and Hans-Joachim Elmers },
title = {Observation of time-reversal symmetry breaking in the band structure of altermagnetic \textrm{RuO$_2$}},
journal = {Science Advances},
volume = {10},
number = {5},
pages = {eadj4883},
year = {2024},
doi = {10.1126/sciadv.adj4883},
eprint = {https://www.science.org/doi/pdf/10.1126/sciadv.adj4883}
}

@article{liu2024absence,
  title = {Absence of Altermagnetic Spin Splitting Character in Rutile Oxide \textrm{RuO$_2$}},
  author = {Liu, Jiayu and Zhan, Jie and Li, Tongrui and Liu, Jishan and Cheng, Shufan and Shi, Yuming and Deng, Liwei and Zhang, Meng and Li, Chihao and Ding, Jianyang and Jiang, Qi and Ye, Mao and Liu, Zhengtai and Jiang, Zhicheng and Wang, Siyu and Li, Qian and Xie, Yanwu and Wang, Yilin and Qiao, Shan and Wen, Jinsheng and Sun, Yan and Shen, Dawei},
  journal = {Phys. Rev. Lett.},
  volume = {133},
  issue = {17},
  pages = {176401},
  numpages = {7},
  year = {2024},
  month = {Oct},
  publisher = {American Physical Society},
  doi = {10.1103/PhysRevLett.133.176401}
}

@article{Hayami2019Nov,
  author = {Hayami, Satoru and Yanagi, Yuki and Kusunose, Hiroaki},
  title = {{Momentum-Dependent Spin Splitting by Collinear Antiferromagnetic Ordering}},
  journal = {J. Phys. Soc. Jpn.},
  volume = {88},
  number = {12},
  pages = {123702},
  year = {2019},
  month = nov,
  issn = {0031-9015},
  publisher = {The Physical Society of Japan},
  doi = {10.7566/JPSJ.88.123702},
}

@article{yuan2020giant,
  title = {Giant momentum-dependent spin splitting in centrosymmetric low-$Z$ antiferromagnets},
  author = {Yuan, Lin-Ding and Wang, Zhi and Luo, Jun-Wei and Rashba, Emmanuel I. and Zunger, Alex},
  journal = {Phys. Rev. B},
  volume = {102},
  issue = {1},
  pages = {014422},
  numpages = {13},
  year = {2020},
  month = {Jul},
  publisher = {American Physical Society},
  doi = {10.1103/PhysRevB.102.014422}
}

@article{mazin2021prediction,
author = {Igor I. Mazin  and Klaus Koepernik  and Michelle D. Johannes  and Rafael González-Hernández  and Libor Šmejkal },
title = {Prediction of unconventional magnetism in doped \textrm{FeSb$_2$}},
journal = {Proceedings of the National Academy of Sciences},
volume = {118},
number = {42},
pages = {e2108924118},
year = {2021},
doi = {10.1073/pnas.2108924118},
eprint = {https://www.pnas.org/doi/pdf/10.1073/pnas.2108924118}
}

@article{jungwirth2026altermagnetic,
  title={Altermagnetic spintronics},
  author={Jungwirth, Tomas and Sinova, J and Wadley, P and Kriegner, D and Reichlova, H and Krizek, F and Ohno, H and {\v{S}}mejkal, L},
  journal={Nature Physics},
  pages={1--10},
  year={2026},
  publisher={Nature Publishing Group UK London}
}

\clearpage
\section*{Extended Data}

\setcounter{figure}{0}
\renewcommand{\thefigure}{\arabic{figure}}
\renewcommand{\figurename}{Extended Data Fig.}

\begin{figure*}[!h]
\begin{center}
\includegraphics[scale=0.76]{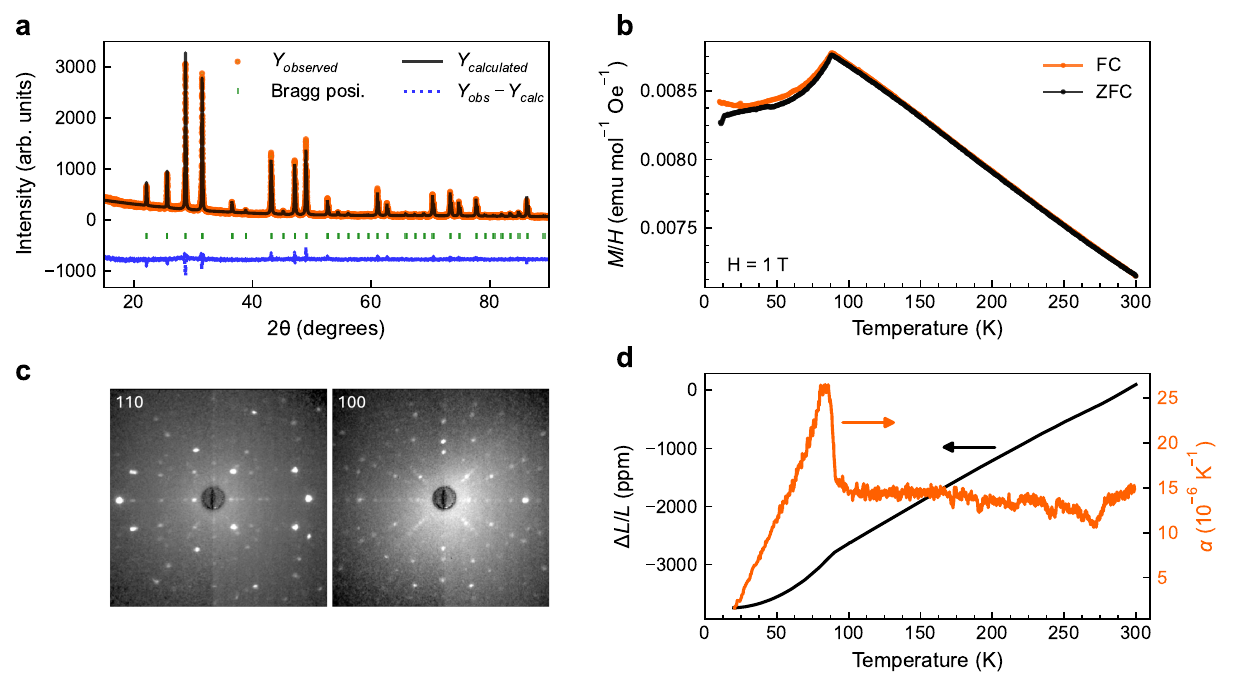}
\caption{ \textbf{Structural and magnetic characterization of MnTe$_2$.} \textbf{a}, Powder XRD data and Rietveld refinement. \textbf{b}, Field-cooled (FC) and zero-field-cooled (ZFC) magnetization measurements performed on a $\langle100\rangle$ oriented single crystal in 1 T field confirm the transition temperature of $T_N \approx$ 87 K. \textbf{c}, Laue back reflection spectra for $\langle110\rangle$ and $\langle100\rangle$ orientations used in dilatometry measurements. \textbf{d}, Thermal expansion of a $\langle100\rangle$ oriented sample. MnTe$_2$ undergoes a spontaneous magnetism-induced lattice contraction upon below $T_N$. This is clearly visible in the thermal expansion coefficient, which is constant at higher temperatures and exhibits an abrupt anomaly at $T_N$. The overall magnitude of the measured thermal expansion agrees with ref. \cite{kasai1981thermal}.  \label{SElaue}}

\end{center}
\end{figure*}

\begin{figure*}[]
\begin{center}
\includegraphics[scale=0.73]{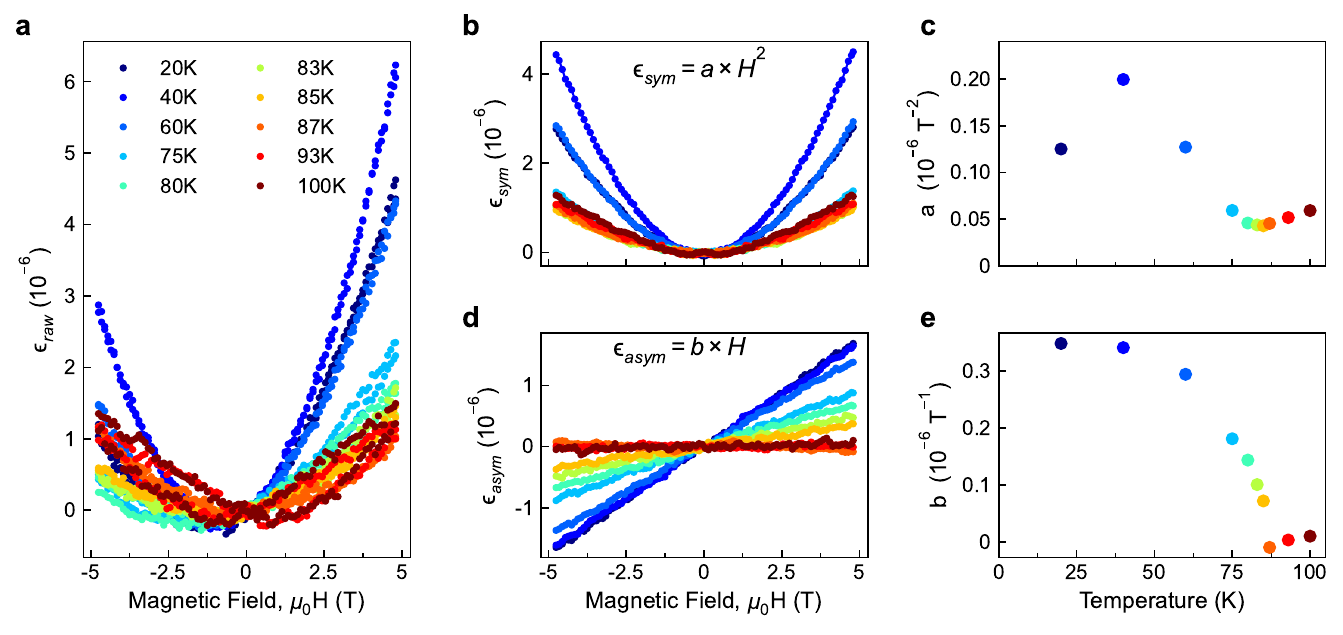}
\caption{\textbf{Transverse magnetostriction measurements for magnetic field H along [001] and $\Delta L$ along [100]}. \textbf{a}, Magnetic field dependence of the relative deformation. The response is primarily quadratic with a small linear component. \textbf{b}, The symmetric component of the relative deformation is quadratic with respect to the magnetic field. \textbf{c}, Temperature dependence of the coefficient of quadratic fit to the symmetric component $a=\partial\epsilon^2_{sym}/2\mu_0\partial H^2$ .\label{100} \textbf{d}, The antisymmetric component of the relative deformation is linear with respect to the magnetic field.  This is likely due to a small misalignment of the crystal in the dilatometer.  \textbf{e} Temperature dependence of the coefficient of linear fit to the antisymmetric component, $b = \partial\epsilon_{asym}/\mu_0\partial H$, plotted as a function of temperature. Its magnitude is close to 14\% of that observed along the [110] direction, suggesting a crystal misalignment of approximately 8\% in the dilatometer. However, given that the maximum misalignment determined from the Laue measurements is less than 3$^\circ$, the larger-than-expected value of $b$ is more likely due to a stronger domain imbalance (compared to [110] measurements) coupling with a smaller misalignment.}

\end{center}
\end{figure*}

\begin{figure*}[]
\begin{center}
\includegraphics[scale=0.7]{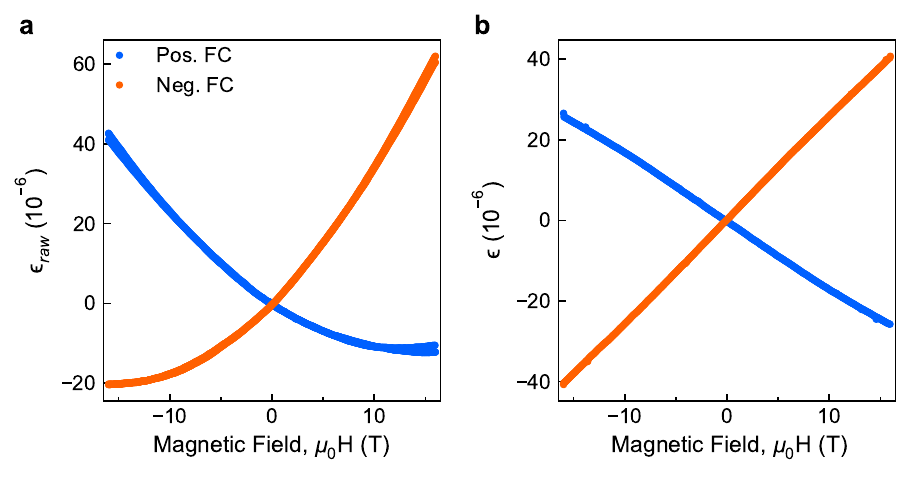}
\caption{ \textbf{Effects of magnetic-field training on magnetostriction.} Transverse magnetostriction measurements for \textbf{H} $\parallel$ [001] and $\Delta \bf L$ $\parallel$ [110], when cooling the sample through the transition in positive (pos. FC) vs. negative (neg. FC) field at 20 K. \textbf{a}, Magnetic field dependence of the relative deformation. The response is primarily linear with a small linear component. \textbf{b}, The antisymmetric component of the relative deformation is perfectly linear with respect to the magnetic field.   \label{16Train}}

\end{center}
\end{figure*}

\begin{figure*}[]
\begin{center}
\includegraphics[scale=0.5]{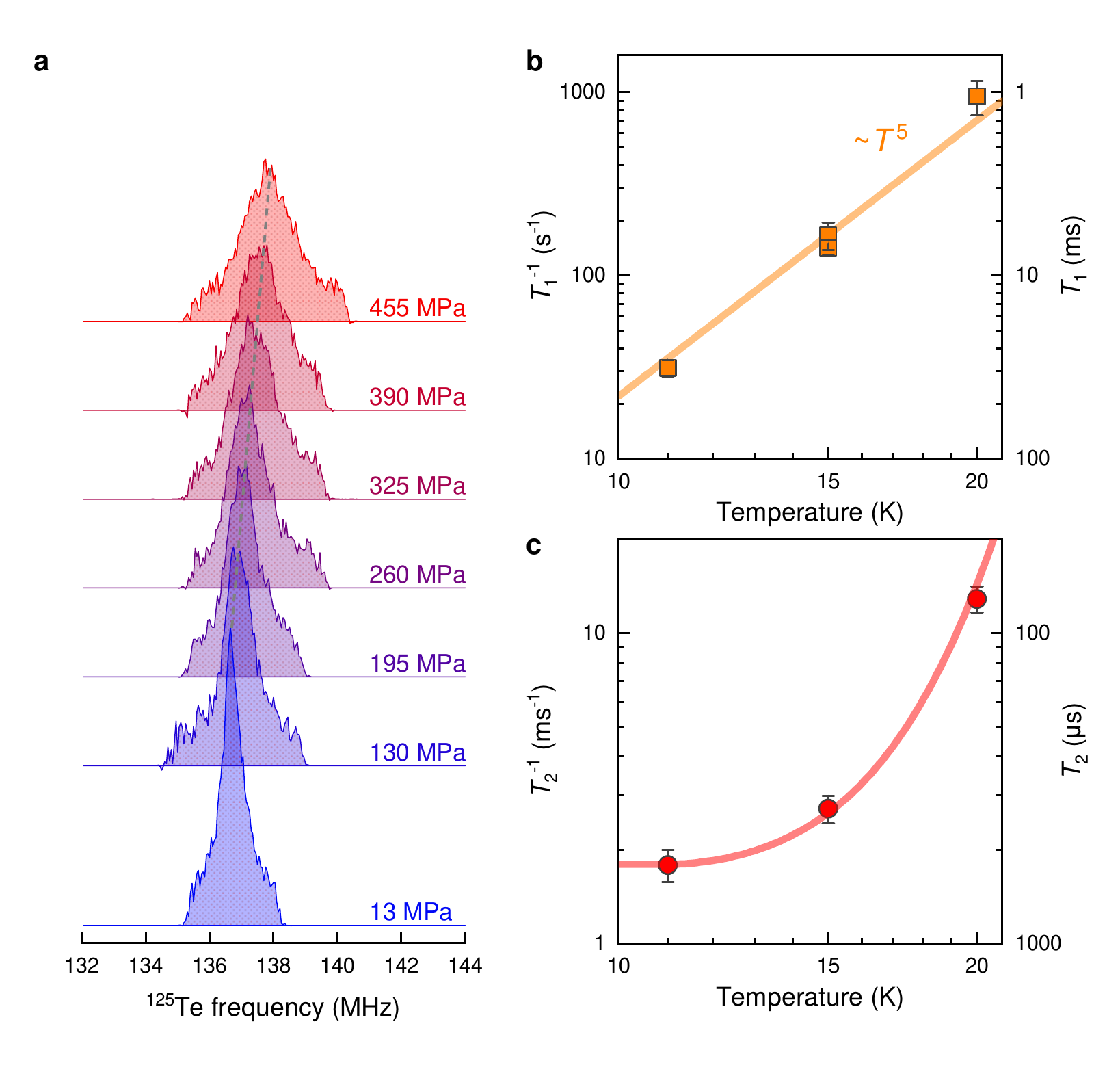}
\caption{\textbf{Additional NMR data}. \textbf{a}, $^{125}$Te spectra for stress along [001], which leads to a shift and broadening of the line, but without splitting. \textbf{b}, $^{125}$Te spin-lattice relaxation rate ($T_1^{-1}$) dependence on temperature at zero stress, showing an extremely fast decrease with cooling, consistent with a $1/T_1 \sim T^5$ dependence (line). \textbf{c}, Observed increase in spin-spin relaxation rate ($T_2^{-1}$) at $T > 20$~K leads to $T_2$-driven signal wipeout well below $T_N \approx 87$ K. The line is a guide to the eye. \label{MoreNMR}}

\end{center}
\end{figure*}

\end{document}